\documentclass[%
 reprint,
 amsmath,amssymb,
 aps,
]{revtex4-1}

\usepackage{graphicx}
\usepackage{dcolumn}
\usepackage{bm}
\usepackage{dsfont}
\usepackage{appendix}

\usepackage{color}
\usepackage{ulem}

\begin{document}

\preprint{APS/123-QED}

\title{Note about the spin connection in general relativity}

\author{Renata Jora
	$^{\it \bf a}$~\footnote[1]{Email:
		rjora@theory.nipne.ro}}

\email[ ]{rjora@theory.nipne.ro}

\affiliation{$^{\bf \it a}$ National Institute of Physics and Nuclear Engineering PO Box MG-6, Bucharest-Magurele, Romania}

\begin{abstract}
In general relativity the fermions are treated from the perspective of the gauged Lorentz group and by introducing the corresponding gauge fields the spin connection.  This procedure is intimately related to the so-called "vielbein" formalism  and stems from the general structure that can be associated to a manifold, the affine connection. In this work we derive the correct spin connection based only on the general covariance of the theory, on the gauged Lorentz symmetry and on the known space-time properties of fermion bilinears generalized to the curved space. Our result coincides exactly with the spin connection obtain through the tetrad formalism.
\end{abstract}
\maketitle

In general relativity the ordinary derivative of a tensor, in order to obtain the general behavior of a tensor, is replaced by the covariant derivative written in terms of an affine connection. For example for a vector the covariant derivative is given by:
\begin{eqnarray}
\partial_{\mu}V_{\nu}\rightarrow D_{\mu}V_{\nu}=\partial_{\mu}V_{\nu}-\Gamma^{\rho}_{\mu\nu}V_{\rho}.
\label{covderibert4553}
\end{eqnarray}
Here $\Gamma^{\rho}_{\mu\nu}$  represents the affine connection which may be independent of a metric. In particular however it is always more amenable to consider an  affine connection that satisfies two main requirements \cite{Caroll}:

a) to be torsion free, i.e. $\Gamma^{\rho}_{\mu\nu}=\Gamma^{\rho}_{\nu\mu}$.

b) to be metric compatible which amount to asking that the covariant derivative of the metric tensor is zero:
\begin{eqnarray}
\Delta_{\rho}g_{\mu\nu}=\partial_{\rho}g_{\mu\nu}-\Gamma^{\sigma}_{\rho\mu}g_{\sigma\nu}-\Gamma^{\sigma}_{\rho\nu}g_{\sigma\mu}=0.
\label{covderibmetric75664}
\end{eqnarray}.

Having established how a derivative of a tensor field must be modified in curved space time one needs to consider another type of fields of relevance in a quantum field theory, the fermion fields. In QFT the fermions lie in a four dimensional representation of the Lorentz group $SO(3,1)$ given by the gamma matrices which span a Clifford algebra with the anti-commutation rule:
\begin{eqnarray}
\{\gamma^a,\gamma^b\}=2\eta^{ab},
\label{antocommut657}
\end{eqnarray}
where $\eta^{ab}$ is the Minkowski metric.  The natural approach in the presence of a curved space-time and of a general coordinate transformation would be then to gauge the Lorentz group and to introduce the gauge fields associated to this, the spin connection. Then in a formalism introduced by Cartan and developed further in \cite{Weyl}, \cite{Isham}  one defines the gamma matrices in the curved space as:
\begin{eqnarray}
\gamma^{\mu}(x)=\gamma^ae^{\mu}_a,
\label{gamma6477}
\end{eqnarray}
where $\gamma^{\mu}$ depend on the coordinate, $\mu$ is the index in the curved space and $a$ is the index in the flat space. The quantities $e^{\mu}_a$ are called a tetrad and satisfy the relation:
\begin{eqnarray}
g^{\mu\nu}=e^{\mu}_ae^{\nu}_b\eta^{ab}.
\label{tetrte5466}
\end{eqnarray}
It is considered that the gamma matrices in the curved space satisfy a generalized Clifford algebra with the anti commutation relation:
\begin{eqnarray}
\{\gamma^{\mu},\gamma^{\nu}\}=2g^{\mu\nu}.
\label{acurvedpsace435}
\end{eqnarray}
If one requires further that the operations of parallel transport and projection on flat and curved indices commute one arrives of the vielbein postulate:
\begin{eqnarray}
D_{\rho}e^m_{\mu}=\partial_{\rho}e^m_{\mu}(x)-\Gamma^{\nu}_{\rho\mu}e^m_{\nu}-\omega^m_{\rho n}e^n_{mu}=0,
\label{vilebepost56477}
\end{eqnarray}
where $\omega^m_{\rho n}$ is the spin  connection which can be extracted from Eq. (\ref{vilebepost56477}) to be:
\begin{eqnarray}
&&\omega_{\mu}^{mn}=e^m_{\nu}\Gamma^{\nu}_{\sigma \mu}e^{\sigma n}+e^m_{\nu}\partial_{\mu}e^{\nu n}=
\nonumber\\
&&e^m_{\nu}\Gamma^{\nu}_{\sigma \mu}e^{\sigma n}-\partial_{\mu}e^m_{\nu}e^{\nu n}.
\label{resspinconn56647}
\end{eqnarray}
Then the covariant derivative of a Dirac fermion in the curved space time is written as:
\begin{eqnarray}
D_{\rho}\Psi=\partial_{\rho}\Psi-\frac{i}{4}\omega_{\rho}^{ab}\sigma_{ab}\Psi,
\label{dreivcov5466377}
\end{eqnarray}
where $\sigma_{ab}=\frac{i}{2}[\gamma^a,\gamma^b]$.

Various attempts have been made in the literature \cite{Q} to introduce fermion covariant derivative  without the use of the vielbein formalism in terms of only the curvilinear coordinates. These involved usually new and complicated mathematical structures and an entire formalism of their own.

In the following we will derive the exact expression for the fermion covariant derivative without the use of the vielbein formalism  by only making some natural assumptions.

Consider two Dirac fermions $\Psi$ and $\bar{\Psi}$ and the gamma matrices $\gamma^{\mu}$ in the curved space time. The main assumption is that the quantity $\bar{\Psi}\gamma^{\mu}\Psi$ transforms as a vector in the curved space-time. Then the quantity,
\begin{eqnarray}
 D_{\rho}[\bar{\Psi}\gamma^{\mu}\Psi],
\label{tens7756}
\end{eqnarray}
where $D_{\rho}$ was introduced in Eq. (\ref{covderibert4553}) should transform as a rank two tensor.

We are interested in writing a covariant derivative for the fermion fields such that the quantity in Eq. (\ref{tens7756})  transforms  as a second rank tensor. Consider that this covariant derivative is expressed as,
\begin{eqnarray}
D_{\rho}\Psi=\partial_{\rho}\Psi+X_{\rho}\Psi,
\label{writtencov455}
\end{eqnarray}
where $X_{\rho}$  may contain in it gamma matrices in the curved space. 

We make the assumption that the covariant derivative operator is linear, although acting with different expression on various fields. Consider the quantity:
\begin{eqnarray}
\bar{\Psi}\Psi Y^{\mu},
\label{res546354}
\end{eqnarray}
where $Y^{\mu}$ is a vector in the curved space. Then,
\begin{eqnarray}
D_{\rho}[\bar{\Psi}\Psi Y^{\mu}],
\label{result87}
\end{eqnarray}
is a rank two tensor. The linearity of the operator implies that the quantity,
\begin{eqnarray}
[D_{\rho}(\bar{\Psi}\Psi)]Y^{\mu}+\bar{\Psi}\Psi D_{\rho}Y^{\mu},
\label{rescs5464}
\end{eqnarray}
is also a second rank tensor. Further on then,
\begin{eqnarray}
D_{\rho}[\bar\Psi\Psi]=[D_{\rho}\bar{\Psi}]\Psi+\bar{\Psi}D_{\rho}\Psi,
\label{res6122}
\end{eqnarray}
is also a second rank tensor. Since $\bar{\Psi}\Psi$ is  a scalar the only second rank tensor that one may form from it is:
\begin{eqnarray}
\partial_{\rho}[\bar{\Psi}\Psi].
\label{res43525}
\end{eqnarray}
Eqs. (\ref{writtencov455}), (\ref{res6122}) and (\ref{res43525}) then imply that:
\begin{eqnarray}
D_{\rho}\bar{\Psi}=\partial_{\rho}\bar{\Psi}-\bar{\Psi}X_{\rho}.
\label{finalre456565}
\end{eqnarray}

In order to determine the composition of $X_{\rho}$ in the $16$ dimensional Lorentz space associated to the gamma matrices we take into account the fact that the covariant derivative is a result of gauging the Lorentz symmetry and therefore should be an expansion in $\sigma^{\mu\nu}$. Then without loss of generality one may write:
\begin{eqnarray}
X_{\rho}=-iA_{\rho\alpha\beta}\sigma^{\alpha\beta},
\label{def453664}
\end{eqnarray}

 In summary one has:
\begin{eqnarray}
&&D_{\rho}\Psi=\partial_{\rho}\Psi-iA_{\rho\alpha\beta}\sigma^{\alpha\beta}
\nonumber\\
&&D_{\rho}\bar{\Psi}=\partial_{\rho}\Psi-iA_{\rho\alpha\beta}\sigma^{\beta\alpha}.
\label{rel76885}
\end{eqnarray}

One can expand  Eq. (\ref{tens7756}) which leads to:
\begin{eqnarray}
&&D_{\rho}[\bar{\Psi}\gamma^{\mu}\Psi]=
\nonumber\\
&&(\partial_{\rho}\bar{\Psi})\gamma^{\mu}\Psi+\bar{\Psi}[\partial_{\rho}\gamma^{\mu}]\Psi+
\nonumber\\
&&\bar{\Psi}\gamma^{\mu}\partial_{\rho}\Psi+\Gamma_{\rho\sigma}^{\mu}\bar{\Psi}\gamma^{\sigma}\Psi.
\label{exp768}
\end{eqnarray}

If we introduce the expression in Eq. (\ref{rel76885}) into Eq. (\ref{exp768}) one obtains that the quantity,
\begin{eqnarray}
&&D_{\rho}\bar{\Psi}\gamma^{\mu}\Psi+\bar{\Psi}\gamma^{\mu}D_{\rho}\Psi+
\nonumber\\
&&+\bar{\Psi}i\gamma^{\mu}A_{\rho\alpha\beta}\sigma^{\alpha\beta}\Psi-\bar{\Psi}iA_{\rho\alpha\beta}\sigma^{\alpha\beta}\gamma^{\mu}\Psi+
\nonumber\\
&&\bar{\Psi}(\partial_{\rho}\gamma^{\mu})\Psi+\Gamma^{\mu}_{\rho\sigma}\bar{\Psi}\gamma^{\sigma}\Psi,
\label{final65774}
\end{eqnarray}
behaves as a second rank tensor. Since the term in the first line of Eq. (\ref{final65774}) behaves like a tensor then also the terms on the second plus the third line must behave as a second rank tensor. Then,
\begin{eqnarray}
\bar{\Psi}i\gamma^{\mu}A_{\rho\alpha\beta}\sigma^{\alpha\beta}\Psi-\bar{\Psi}iA_{\rho\alpha\beta}\sigma^{\alpha\beta}\gamma^{\mu}\Psi+
\nonumber\\
\bar{\Psi}(\partial_{\rho}\gamma^{\mu})\Psi+\Gamma^{\mu}_{\rho\sigma}\bar{\Psi}\gamma^{\sigma}\Psi=T_{\rho}^{\mu},
\label{final65774568}
\end{eqnarray}
where $T^{\mu}_{\rho}$ is an arbitrary tensor expressed in terms of the fermion fields. Since there is not such tensor besides those introduced at this point with the correct mass dimension one can consider this tensor zero.

One may rewrite Eq. (\ref{final65774568}) as:
\begin{eqnarray}
iA_{\rho\alpha\beta}[\gamma^{\mu},\sigma^{\alpha\beta}]=-\Gamma^{\mu}_{\rho\sigma}\gamma^{\sigma}-\partial_{\rho}\gamma^{\mu}.
\label{finalsecrel756488}
\end{eqnarray}

In the flat space we know that:
\begin{eqnarray}
\frac{1}{2}[\gamma^a,\sigma^{bc}]=i(\eta^{ab}\gamma^c-\eta^{ac}\gamma^b).
\label{res6455344}
\end{eqnarray}
Since we consider in the curved space a similar Clifford algebra this time with the gamma matrices space time dependent the same relation should work if the flat indices would be replaced by the curved indices. Then Eq. (\ref{finalsecrel756488}) becomes:
\begin{eqnarray}
-4A_{\rho\,\beta}^{\mu}\gamma^{\beta}=-\Gamma^{\mu}_{\rho\sigma}\gamma^{\sigma}-\partial_{\rho}\gamma^{\mu}.
\label{rel6775}
\end{eqnarray}
We multiply Eq. (\ref{rel6775}) by $\gamma^{\lambda}$ and take the trace to obtain:
\begin{eqnarray}
A_{\rho\,\lambda}^{\mu}=\frac{1}{4}\Gamma^{\mu}_{\rho\lambda}+\frac{1}{16}{\rm Tr}[\gamma^{\lambda}\partial_{\rho}\gamma^{\mu}].
\label{oneofthefinal75664}
\end{eqnarray}
Finally the covariant derivative for the fermion fields is written in terms of only quantities in the curved  space as:
\begin{eqnarray}
D_{\rho}\Psi=\partial_{\rho}\Psi-iA_{\rho\,\beta}^{\alpha}\sigma_{\alpha}^{\,\beta}\Psi,
\label{finalrelfinal65774}
\end{eqnarray}
where $A_{\rho\,\beta}^{\alpha}$ is given in Eq. (\ref{oneofthefinal75664}).

Next we will show that the spin connection introduced in Eq. (\ref{finalrelfinal65774}) is identical to that in Eq. (\ref{dreivcov5466377}). For that we write:
\begin{eqnarray}
&&-iA_{\rho\,\lambda}^{\mu}\sigma_{\mu}^{\,\lambda}=
\nonumber\\
&&-i\frac{1}{4}[\Gamma^{\mu}_{\rho\lambda}+\frac{1}{4}{\rm Tr}[\gamma^{\lambda}\partial_{\rho}\gamma^{\mu}]]\sigma_{\mu}^{\,\lambda}=
\nonumber\\
&&-i\frac{1}{4}[\Gamma^{\mu}_{\rho\lambda}+\partial_{\rho}e^{\mu}_ae^{\lambda}_b\eta^{ab}]e_{\mu c}e^{\lambda}_ d\sigma^{cd}=
\nonumber\\
&&-i\frac{1}{4}[\Gamma^{\mu}_{\rho\lambda}e_{\mu c}e^{\lambda}_ d+\partial_{\rho}e^{\mu}_de_{\mu c}]\sigma^{cd}.
\label{resfinal5677676}
\end{eqnarray}

Eq. (\ref{resfinal5677676}) shows that the exact expression of the spin connection obtained through the vielbein formalism can be obtained by using only quantities defined in the curved space with the gamma matrices in the curved space satisfying a similar Clifford algebra.


\begin{thebibliography}{30}

\bibitem{Caroll}  S. Caroll, arXiv:gr-qc/9712019 (1997).
\bibitem{Weyl} H. Weyl, Z. Phys. {\bf 56}, 330 (1929); T. W. Kibble, J. Math. Phys. {\bf 2}, 212 (1961).
\bibitem{Isham} S. Deser and C. J. Isham, Phys. Rev. D {\bf 14}, 2505 (1976).
\bibitem{Q} A. Weldon, Phys. Rev. D {\bf 63}, 104010 (2010).














\end{thebibliography}
\end{document}